\documentclass[a4paper,11pt]{article}
\usepackage{graphicx}
\usepackage{multicol}
\usepackage{subfigure}
\usepackage{graphicx,epsfig}
\usepackage{epstopdf}
\usepackage{latexsym}
\usepackage{amsfonts,amsmath,amssymb}
\usepackage{hyperref}
\usepackage{float}

\usepackage{epsfig,multicol,bbm}

\newcommand\fverb{\setbox\pippobox=\hbox\bgroup\verb}
\newcommand\fverbit{\egroup\item[\fbox{\unhbox\pippobox}]}

\newbox\pippobox

\begin{document}
\title{\bf Boosted Kaluza-Klein Magnetic Monopole }
\author{S. Sedigheh Hashemi   \,\,and \,\,  Nematollah Riazi\thanks{Electronic address: n\_riazi@sbu.ac.ir (Corresponding author).} 
\\
\small Department of Physics, Shahid Beheshti University, G.C., Evin, Tehran 19839,  Iran}
\maketitle
\begin{abstract}
We consider a Kaluza-Klein vacuum solution which is  closely related to 
the Gross-Perry-Sorkin (GPS) magnetic monopole. The solution  can be  obtained from the Euclidean Taub-NUT solution with an extra compact fifth spatial dimension within the formalism of Kaluza-Klein reduction. We study its physical
properties as appearing in $(3+1)$ spacetime dimensions, which turns out to be  a static magnetic monopole. We then boost the GPS magnetic monopole along the extra dimension, and perform the Kaluza-Klein reduction. The resulting  four-dimensional spacetime is a rotating stationary system, with both electric and magnetic fields. In fact, after the boost the magnetic monopole 
turns into a string connected to a dyon.
\end{abstract}

\section{Introduction: Kaluza-Klein theory}\label{sec1}
The possibility that the universe is embedded in a $(4+d) $ dimensional world has gained the attention of many researches.
In the Randall and Sundrum \cite{RS} theory the  matter and fields are restricted to a four dimensional space-time known as brane which is embedded in a five dimensional spacetime (bulk). In Space-Time-Matter (STM) theory \cite{stm} all physical quantities such as matter density and pressure, gain a  geometrical interpretation.
Among these  higher dimensional theories, the original  Kaluza-Klein theory  unifies gravity and electromagnetism \cite{1}, assuming that the fifth dimension is compact \cite{5}.

Kaluza\rq{}s idea was that the universe has four spatial
dimensions, and the extra dimension is compactified  to form a
circle so small as to be unobservable \cite{17}. Klein's contribution was to
make a reasonable physical basis for the  compactification  of the
fifth dimension \cite{13}, \cite{18}. This school of thinking later led to
the eleven-dimensional supergravity theories in  1980s and to the
\lq\lq{}theory of everything\rq\rq{} or ten-dimensional
superstring theory \cite{14}.

With this  unification
 the vacuum $(4+1)$ dimensional Kaluza-Klein solutions that is $\hat G_{AB}=0$ (the
indices $A,B,...$ run over $0...4$ ), will reduce to the $(3+1)$ Einstein field equations with
effective matter and the curvature in $(4+1)$ spacetime induces matter
in $(3+1)$ dimensional spacetime \cite{16}. In the context of Kaluza-Klein, the Einstein tensor has the usual definition $\hat G_{AB} \equiv \hat
R_{AB}-1/2\hat R \hat g_{AB}$, where $\hat R_{AB}$ and $\hat R = \hat g_{AB} \hat R^{AB}$ are  the
five-dimensional Ricci tensor and scalar,  respectively, and $\hat
g_{AB}$ is the metric tensor in  five dimensions \cite{14}. The $\mu \nu$ part  of $\hat g^{AB}$, which is  $g^{\mu \nu}$
is the contravariant four dimensional
metric tensor, and the electromagnetic potential and the scalar field are given by $A^{\mu}$  and $\phi$, respectively.
 The general correspondence  between the above
components is given by 
\begin{equation}\label{eq1}
\hat g^{AB}= \left(
‎\begin{array}{ccccccc}‎
‎g^{\mu \nu}‎   & -\kappa A^{\mu} &  \\‎
 ‎‎  &‎\\‎
-\kappa A^{\nu}    ~~& \kappa^2 A^{\sigma}A_{\sigma}+\phi^2 &  \\‎
‎\end{array}
‎\right),
\end{equation}
where $\kappa$ is a coupling  constant for  the electromagnetic potential $A^{\mu}$ \cite{8},\cite{Y}.

Many spherically symmetric solutions of Kaluza-Klein  type  are investigated in \cite{2} and \cite{7}.
Among these solutions, the Gross-Perry-Sorkin (GPS) spacetime \cite{5},\cite{26} is one of the exact vacuum solutions of Einstein field equations in five-dimensional gravity  which is  stationary and without event horizon representing a magnetic monopole  which is usually called the Kaluza-Klein monopole\cite{G},\cite{j}.  As it is well known, the theory of magnetic monopole was formulated by Dirac in $1931$ \cite{Dirac}. He showed that the electric charge quantization can be explained by  the existence of a magnetic monopole. In addition to magnetic charge, monopoles are characterized by their peculiar topology. They carry one unit of Euler character, and consequently one can construct stationary dipole solutions from them \cite{M}.

The Kaluza-Klein monopole has an important role in M/String theory. As an example, a Ricci-flat eleven dimensional Lorentzian metric can be obtained from Kaluza-Klein monopole metric times six flat Euclidean dimensions, which when reduced to ten dimensional spacetime can be interpreted as a {\rm}D$6$ brane solution of {\rm } II{\rm}A  string theory \cite{G}.

In this paper, we consider a  vacuum solution of Kaluza-Klein
theory in five-dimensional spacetime which is closely related to
the Taub-NUT  and GPS metric. 
The Taub-NUT  solution has many interesting
features; it carries a particular type of charge (NUT charge), which has
topological origins and can be regarded as \lq\lq{}gravitational
magnetic charge\rq\rq{}\cite{ortin}, \cite{UK}. We boost the  magnetic monopole along the fifth dimension  and investigate its properties in the four-dimensional spacetime by using the Kaluza-Klein reduction. The monopole turns into a dyon connected to a magnetically charged string.

The plan of this paper is as follows. In section \ref{3},  we 
will  review a Taub-NUT-like Kaluza-Klein solution and
investigate its physical properties in four dimensions.  In section \ref{4}, we will study the boosted Kaluza-Klein magnetic monopole, and explore its physical properties.
 In the last section we will draw our main conclusions.


\section{  Kaluza-Klein Magnetic Monopole and Taub-NUT Solution}\label{3}
\textbf{In this section, we first review the main features of the Kaluza-Klein monopole before the boost.}
The Kaluza-Klein monopole, known also as Gross-Perry-Sorkin solution, is a generalization of the self-dual Euclidean Taub-NUT solution. The Taub-NUT solution was first discovered by Taub (1951), and
subsequently by Newman, Tamburino and Unti (1963) as a
generalization of the Schwarzschild spacetime \cite{21}, \cite{22}. This
solution is a single,  non-radiating  and  analytic extension of the Taub universe, the anisotropic but spatially homogeneous
vacuum solution  of
Einstein field equations with topology $R^1 \times S^3$.
The Taub metric is given by
\begin{equation}
{\rm d}s^2=-\frac{1}{V(t)}{\rm d}t^2+4b^2V(t)({\rm d}\psi+\cos \theta {\rm d}\phi)^2+(t^2+b^2)({\rm d}\theta^2+\sin \theta^2{\rm d}\phi^2),
\end{equation}
where $V(t)=-1+ 2(mt+b^2)(t^2+b^2)^{-1}$,  $m$ and $b$ are positive constants, $\psi, \phi, \theta$ are Euler angels with usual ranges \cite{k}.
The Taub-NUT solution is nowadays
being involved in the context of higher-dimensional theories of
semi-classical quantum gravity \cite{20}. As an  example, in the
work by Gross and Perry \cite{5} and  Sorkin \cite{4},  soliton
solutions were obtained by embedding the Taub-NUT gravitational
instanton inside the five dimensional Kaluza-Klein
manifold \cite{21}.  One  such solution which obeys the Dirac quantization
condition is considered in \cite{5}.

The Kaluza-Klein monopole of Gross-Perry-Sorkin  is  represented by the
following metric \cite{5}
\begin{align}\label{eq4}
 {\rm d}s^2=- {\rm d}t^2+V( {\rm d}x^5+4m(1-\cos\theta) {\rm d}\phi)^2+\frac{1}{V}( {\rm d}r^2+r^2 {\rm d} \theta^2 +r^2 \sin ^2\theta  {\rm d}\phi^2),
\end{align}
where
\begin{align}
\frac{1}{V}=&1+\frac{4m}{r}.
\end{align}
The Taub-NUT instanton is obtained by putting ${\rm d}t=0$.
For this solution the coordinate singularity is located at $r=0$, which is called   NUT singularity. This can be vanished if the extra coordinate  $x^5$ is periodic with period $16\pi m=2\pi R$, where $R$ is the radius
of the fifth dimension. Thus   $m=\sqrt{\pi G}/2e$ \cite{28}.
The gauge field $A_{\nu}$ is given by $A_{\phi}=4m(1-cos\theta)$,
and the magnetic field is $B=4m\bold{r}/r^3$, which is
clearly that of a monopole and has a Dirac string singularity in
the range $r=0 $ to $\infty$. The magnetic charge of this monopole
is $g=m/\sqrt{\pi G}$
 which has one unit of Dirac charge. In this model,
 the total magnetic flux is constant.
For this solution, the soliton mass is determined to be
$M^2=m_{p}^2/16\alpha$ where $m_{p}$ is the Planck mass and $\alpha$ is the fine-structure constant.

In our previous work \cite{ssh}, we   presented a metric which is a vacuum five dimensional
solution, having some properties in common with the monopole of
Gross-Perry-Sorkin, despite some differences. In this part, we will  briefly review the results we obtained there (see \cite{ssh}), and will then extend the results in coming sections. The metric is given by
\begin{align}\label{55}
 {\rm d}s^2_{(5)}= - {\rm d}t^2+(1-\frac{2m}{r})\left( {\rm d}r^2+r^2 {\rm d}\theta^2 +r^2\sin^2 \theta  {\rm d}\phi^2\right)+
\left(\frac{4m^2}{1-\frac{2m}{r}}\right)\left( {\rm d}\psi + \cos\theta  {\rm d}\phi\right)^2,
\end{align}
where, the extra coordinate is represented by $\psi$ \footnote{Note that not only the sign of $m$ is taken differently from (\ref{eq4}), but also the structure of the metric is different, leading to some essentially different results.}. The coordinates take on the usual ranges
$r\geq0$, $0\leq\theta\leq\pi$, $0\leq  \phi\leq 2\pi$ and $0\leq
\psi \leq 2\pi$. 
It should be noted that the  metric (\ref{55}) can be obtained from (\ref{eq4}) by replacing ${\rm d}x^5=2m\left({\rm d}\psi + {\rm d}\phi\right)$ and $m \rightarrow -m/2$.
It should be noted that, for negative $m$, we will still have a vacuum solution. In the next section, we will consider this case for some of our results.

The Killing vectors associated with metric (\ref{55})   are given
by
\begin{align}
&K_{0}=(1,0,0,0,0),\quad  K_{1}=(0,0,0,0,1), \quad  K_{2}=(0,0,0,1,0),\nonumber\\
&K_{3}=(0,0, -\sin\phi, -\cot\theta \cos \phi, \csc \theta \cos\phi),\nonumber\\
&K_{4}=(0,0, \cos\phi, -\cot\theta \sin\phi, \csc \theta \sin\phi),
\end{align}
which are  the same as in the Taub-NUT metric  discussed in
\cite{25}, where the authors studied spinning particles in the
Taub-NUT space.

  The gauge field $A_{\mu}$,  and the scalar field
$\phi$ deduced from the  metric  (\ref{14}) with the help of (\ref{eq1}) are
$A_{\phi}=cos\theta/\kappa$, and $ \phi^2=4m^2/(1-\frac{2m}{r})$,
respectively. Moreover, the  electromagnetic tensor is $F_{\theta \phi}=-F_{\phi \theta}= -\sin \theta/\kappa$,
which corresponds to a radial magnetic field $B_{r}=1/\kappa r^2$ with a magnetic  
charge $Q_{M}=1/\kappa$.  The total magnetic flux through any spherical surface centered at
the origin  can be calculated via \cite{26} leading to  the result $\Phi_{B}=2\pi/\kappa$, which  is a constant
(i.e. we have a point-like magnetic charge).

The four dimensional metric deduced  from  Eq.  (\ref{14})  with the use of (\ref{eq1})
leads to the following   asymptotically flat spacetime:
\begin{align}\label{26}
 {\rm d}s^2_{\left(4\right)}=- {\rm d}t^2+\left(1-\frac{2m}{r}\right) {\rm d}r^2+r^2\left(1-\frac{2m}{r}\right)\left( {\rm d}\theta^2 +\sin ^2\theta 
 {\rm d}\phi^2\right).
\end{align}
 The four dimensional metric (\ref{26}) has two curvature singularities at $r=0$ and $r=2m$ unless $m<0$.
If we calculate the  surface area of a $S^2$ hypersurface at constant $t$
and $r$, we see that the surface area $A\left(r\right)$ becomes
zero at $r=0$, as well as $r=2m$. This means that the $r=2m$
  hypersurface is in fact  a point (i.e. a sphere with zero surface area).
For  $r>2m$ the signature of the metric is proper $(-,+,+,+)$
but for the range $0<r<2m$ the signature of
 the metric will be improper
and non-Lorentzian $ (-,-,-,-)$, thus the patch
$r<2m$ is excluded from the physical spacetime. therefore,  this spacetime is
 considered only in the range  $r\geq2m$.
 Since the range $0<r<2m$ is removed from the spacetime, there remains only one curvature singularity at
 $r=2m$.

By computing the  components of the energy-momentum tensor for the metric (\ref{26}),
one can  show that the effective matter field around the
singularity can not be considered as an ultra-relativistic quantum
field (or radiation) in contrast to the Kaluza-Klein solitons described in
\cite{23}.  On the other hand, the gravitational mass was derived in two ways
and it was shown to vanish ($M_{g}=0$).

 It turns out that  the Kaluza-Klein monopole in isotropic coordinates gets the asymptotic form
\begin{equation}\label{35}
 {\rm d}s^2=-(1+\frac{2m}{r})^{-1/2} {\rm d}t^2+(1+\frac{2m}{r})^{1/2}( {\rm d}r^2+r^2 {\rm d}\Omega^2).
\end{equation}

 If in the process of compactification from $(4+1)$ to $(3+1)$ dimensions we use the ansatz
\begin{equation}
\hat G_{AB}= \phi^{\beta}\left(
‎\begin{array}{ccccccc}‎
‎g_{\mu \nu}+\phi A_{\mu}A_{\nu}‎   & \phi A_{\mu} &  \\‎
 ‎‎  &‎\\‎
\phi A_{\nu}    ~~& \phi &  \\‎
‎\end{array}
‎\right),
\end{equation}
then the choice of $\beta$ for which $\phi$ does not appear explicitly is called the Einstein frame.
Using the above equation will lead to the four dimensional metric
\begin{align}\label{37}
 {\rm d}s^2_{\left(4\right)}=\phi^{-\beta}\left[- {\rm d}t^2+\left(1-\frac{2m}{r}\right) {\rm d}r^2+r^2\left(1-\frac{2m}{r}\right) {\rm d}\Omega^2\right].
\end{align}
By  choosing $\phi^{-\beta}=(1-\frac{2m}{r})^{-\frac{1}{2}}$, equation (\ref{37}) reduces to
\begin{align}\label{38}
 {\rm d}s^2=-(1-\frac{2m}{r})^{-1/2}  {\rm d}t^2+(1-\frac{2m}{r})^{1/2}( {\rm d}r^2+r^2  {\rm d}\Omega^2),
\end{align}
which is the same as (\ref{35}) if we replace $m$ by $-m$.

\section{The Boosted Kaluza-Klein Magnetic Monopole}\label{4}
In this section, we apply a  boost to the  Kaluza-Klein magnetic monopole which satisfies the vacuum Einstein field equations. The proposed boost is along the extra dimension $\psi$ with the boost parameter $\alpha$. We consider   metric (\ref{55}) with coordinate renamed as  $(t^{\prime}, r^{\prime},\theta ^{\prime}, \phi ^{\prime}, \psi ^{\prime})$, and define the boosted coordinates as $(t, r, \theta, \phi, \psi)$. Then  we apply the following transformations
\begin{eqnarray}
t^{\prime}&= &t\cosh \alpha -\psi \sinh \alpha~, \\
\psi ^{\prime}&= &\psi \cosh \alpha-t \sinh \alpha,
\end{eqnarray}
with the above transformations, the metric  becomes
\begin{align}\label{14}
{\rm d}s^2&=-\left(\cosh ^2\alpha -\frac{4 m^2 \sinh ^2\alpha}{1-\frac{2 m}{r}}\right){\rm d}t^2+(1-\frac{2m}{r})\left({\rm d}r^2+r^2{\rm d}\theta^2\right)
\nonumber\\&+\left(    \frac{4 m^2  \cos ^2\theta}{1-\frac{2 m}{r}}+r^2 (1-\frac{2 m}{r}) \sin ^2\theta   \right){\rm d}\phi^2+\left(-\sinh ^2\alpha +\frac{4 m^2 \cosh ^2\alpha}{1-\frac{2 m}{r}}\right){\rm d}\psi^2
\nonumber\\&+\sinh 2\alpha\left(1-\frac{4 m^2  }{1-\frac{2 m}{r}}\right){\rm d}t{\rm d}\psi+
\frac{8 m^2  }{1-\frac{2 m}{r}}\cos \theta \cosh \alpha {\rm d}\phi {\rm d}\psi-\frac{8m^2}{1-\frac{2m}{r}}\cos \theta \sinh \alpha{\rm d}\phi {\rm d}t.
\end{align}
which is no longer a static solution because of the ${\rm d}\phi {\rm d}t$ term. 
The metric, however, remains stationary (i.e. $\partial {g_{AB}}/\partial {t}=0$).
It should be stressed that for obtaining the main results of the present paper, which appear after the boost, it is not essential to choose a particular sign for $m$ (we will consider both possibilities in what follows). Let us rewrite (\ref{14}) in the form
\begin{align}\label{15}
{\rm d}s^2&=-\left(\cosh ^2\alpha -\frac{4 m^2 \sinh ^2\alpha}{1+\frac{2 m}{r}}\right){\rm d}t^2+(1+\frac{2m}{r})\left({\rm d}r^2+r^2{\rm d}\theta^2\right)
\nonumber\\&+\left(    \frac{4 m^2  \cos ^2\theta}{1+\frac{2 m}{r}}+r^2 (1+\frac{2 m}{r}) \sin ^2\theta   \right){\rm d}\phi^2+\left(-\sinh ^2\alpha +\frac{4 m^2 \cosh ^2\alpha}{1+\frac{2 m}{r}}\right){\rm d}\psi^2
\nonumber\\&+\sinh 2\alpha\left(1-\frac{4 m^2  }{1+\frac{2 m}{r}}\right){\rm d}t{\rm d}\psi+
\frac{8 m^2  }{1+\frac{2 m}{r}}\cos \theta \cosh \alpha {\rm d}\phi {\rm d}\psi-\frac{8m^2}{1+\frac{2m}{r}}\cos \theta \sinh \alpha{\rm d}\phi {\rm d}t.
\end{align}

The transformed scalar and gauge fields from (\ref{14})  are 
\begin{equation}
\phi^2=\frac{4 m^2 \cosh ^2\alpha }{1-\frac{2 m}{r}}-\sinh ^2\alpha,
\end{equation}
\begin{equation}
A_t=-\frac{\left(4 m^2 r+2 m-r\right) \sinh 2 \alpha}{2 \kappa  \left(4 m^2 r \cosh ^2\alpha +(2
   m-r) \sinh ^2\alpha\right)},
\end{equation}
and
\begin{equation}
A_{\phi}=\frac{4 m^2  r \cosh \alpha \cos \theta}{\kappa  \left(4 m^2 r \cosh ^2\alpha+(2
   m-r) \sinh ^2\alpha\right)},
\end{equation}
which lead to the following electromagnetic field components  
\begin{equation}
F_{r\phi}=\frac{8 m^3  \sinh ^2\alpha  \cosh \alpha \cos \theta }{\kappa  \left(4 m^2 r \cosh
   ^2\alpha+(2 m-r) \sinh ^2\alpha\right)^2}=-r\sin \theta B_{\theta},
\end{equation}
\begin{equation}\label{43}
F_{rt}=\frac{4 m^3 \sinh 2 \alpha}{\kappa  \left(4 m^2 r \cosh ^2\alpha+(2 m-r) \sinh ^2\alpha
   \right)^2}=E_r,
\end{equation}
\begin{equation}\label{br}
F_{\theta \phi}=-\frac{4 m^2  r \cosh \alpha \sin \theta}{\kappa  \left(4 m^2 r \cosh ^2\alpha+(2
   m-r) \sinh ^2\alpha\right)}=r^2\sin \theta B_r.
\end{equation}
It can be seen that by setting $\sinh \alpha=0$, the boosted solution will reduce to the previous metric (\ref{55}). The results just obtained are valid for both signs of $m$.

If we perform a Kaluza-Klein reduction, the four dimensional spacetime for the boosted  metric (\ref{14}) becomes
\begin{align}\label{44}
{\rm d}s^2_{(4)}=&-\frac{4 m^2 r}{4 m^2 r \cosh ^2\alpha +(2 m-r) \sinh ^2\alpha}{\rm d}t^2+\left(1-\frac{2m}{r}\right)\left({\rm d}r^2+r^2{\rm d}\theta^2\right)
\nonumber\\&
+\frac{r \left(-\sinh ^2\alpha  \left(4 m^2  \cos ^2\theta+(r-2 m)^2 \sin ^2\theta
  \right)-4 m^2 r (2 m-r) \cosh ^2\alpha \sin ^2\theta\right)}{4 m^2 r \cosh ^2\alpha +(2
   m-r) \sinh ^2\alpha}{\rm d}\phi^2\nonumber\\&-\frac{4 m^2  r \sinh \alpha \cos \theta }{4 m^2 r \cosh ^2\alpha + (2 m-r)\sinh ^2\alpha
}{\rm d}t{\rm d}\phi.
\end{align}

The four dimensional solution (\ref{44})  is singular at 
\begin{equation}\label{24}
r_1=-\frac{2 m \sinh ^2\alpha }{4 m^2 \cosh ^2\alpha -\sinh ^2\alpha},~~~~r_2=0.
\end{equation}
 The Ricci scalar for metric (\ref{44})  can be calculated easily, and   diverges at the following locations in addition to $r_1$,  and $r_2$ 
\begin{eqnarray}
r_3&=&\frac{1}{7} \csc ^2\theta  \left(6 \sin ^2\theta +\sqrt{\sin ^2\theta \left(64-43 \cos ^2\theta \right)}\right)~, \\
r_4&=&\frac{1}{7} \csc ^2\theta \left(6 \sin ^2\theta-\sqrt{\sin ^2\theta \left(64-43 \cos ^2\theta \right)}\right)~, \\
r_5&=&0,~~~ r_6=2,~~~  r_7=-\frac{2}{7}.
\end{eqnarray}
in which $r_3$ to $r_7$ are simplified after  arbitrarily setting 
  $\sinh \alpha=1$, $\cosh \alpha=\sqrt{2}$, and $m=1$.  These  points are curvature singularities since the Ricci scalar, and the nontrivial quadratic curvature invariant $R^{\mu \nu \alpha \beta}R_{\mu \nu \alpha \beta}$ diverge. With these assumptions,  $r_1$ equals $-2/7$, therefore
   $r_7$  and $r_{1}$ are irrelevant since they become negative. It is also easy to see that $r_4$ is   a negative function of coordinate $\theta$. Therefore, only $r_2=r_5=0$, and $r_3(\theta)$ are relevant. It should be noted that the singularity $r_{3}$ can be removed by choosing suitable values for the parameters $m$ and $\alpha$ (e.g. $ m \simeq 5/100$, $\alpha\simeq \pi/6$ lead to an imaginary value for $r_{3}$).
  Also, the singularities for four dimensional metric deduced from (\ref{15}) with arbitrarily setting 
  $\sinh \alpha=1$, $\cosh \alpha=\sqrt{2}$, and $m=1$ are given by
   \begin{eqnarray}
    r\rq{}_1&=&-\frac{1}{7} \csc ^2\theta  \left(6 \sin ^2\theta +\sqrt{\sin ^2\theta \left(64-43 \cos ^2\theta \right)}\right)~, \\
r\rq{}_2&=&\frac{1}{7} \csc ^2\theta \left(-6 \sin ^2\theta+\sqrt{\sin ^2\theta \left(64-43 \cos ^2\theta \right)}\right),\\
  r\rq{}_3 &= &\frac{2}{7}, \quad r\rq{}_4 = 0\quad r\rq{}_5 = -2,
   \end{eqnarray}
in which, $r\rq{}_1$ is a negative function of $\theta$, and $r\rq{}_5$ is also negative, thus not physical.

In order to analyze the nature of the singularities, let us calculate the surface  area of a $S^{2}$ hypersurface of constant $t$ and $r$ for metric (\ref{44})
\begin{equation}
A\left(r\right)=\int \sqrt{\left|g^{\left(2\right)}\right|} {\rm d}x^2=8\pi \sqrt{\left|\frac{(2-r)r^2}{7 r+2}\right|} E\left(\frac{7 r^2}{4}-3 r\right),
\end{equation}
where $E$ stands for the elliptic integral. The result of the surface area is zero for the singularities $r=0$, and $r=2$ (the integrand is simplified after setting  $\sinh \alpha=1$, $\cosh \alpha=\sqrt{2}$),  which means that in our coordinate system $r=0$, and $r=2$ are  points. By using a coordinate transformation $\tilde{r}^2=r^2(1-2m/r)$, these two singularities will  be transformed to  $\tilde{r}=0$. Furthermore, the determinant of the metric  (\ref{44}) 
is positive in the range $0<r<2m$, and negative for the range $r>2m$, thus the range $0<r<2m$ is removed from the spacetime because of having an improper signature.

  The infinite redshift  surface $r_{ir}$ for metric (\ref{44})  can exist if the  condition $r_{ir}>0$ holds, i.e.
\begin{equation}
4 m^2 \cosh ^2\alpha -\sinh ^2\alpha <0 \Rightarrow m^2<\frac{1}{4}\tanh^2 \alpha.
\end{equation}
The Killing horizon can be obtained with the condition $\xi^2=0$, where $\xi$ is a timelike Killing vector, that is  $ \xi^{\mu}=\partial x^{\mu}/\partial t$, therefore we have
\begin{equation}
g_{\mu \nu}\xi^{\mu}\xi^{\nu}=-\frac{4 m^2 r}{4 m^2 r \cosh ^2\alpha +(2 m-r) \sinh ^2\alpha}=0,
\end{equation}
which gives $r=0$. There are therefore no Killing or event horizons in the physical spacetime.
 The event horizon can be obtained by $g^{rr}=0$, which also corresponds to  $r=0$. 

For the four dimensional boosted Kaluza-Klein solution, we infer from (\ref{43}) and (\ref{br}) that the radial electric field $E_{r}$, and the  radial magnetic field $B_{r}$ do not vanish and consequently, one can find the net electric   and  magnetic fluxes through any two-dimensional surface.

The electric flux may be computed via \cite{caroll} 
\begin{align}
Q_{\rm E}=-\int _{\partial \Sigma}{\rm d}^{n-2}z\sqrt{|\gamma^{\partial \Sigma}|}n_{\mu}\sigma_{\nu}F^{\mu \nu},
\end{align}
where $\Sigma$ is a hypersurface which is typically a hypersurface of constant $t$ and $r$,
 $|\gamma^{\partial \Sigma}|$ is the determinant of the  induced metric on the boundary $\partial \Sigma$,  $n$ and $\sigma$ are the unit normal vectors to boundary given by
\begin{align}
n^{\mu}=(1,0,0,0), ~~~\sigma^{\mu}=(0,1,0,0),
\end{align}
 hence
\begin{align}
Q_{\rm E}=\lim_{r\rightarrow \infty }\int_{s^2} E_rn^{t}\sigma^{r}g^{tt}g_{tt}     r^2\sin \theta {\rm d}\theta {\rm d}\phi.
\end{align}
After calculating the integral for metric (\ref{44}), the result will be a function of $r$,  which indicates that the charge is not point-like, but extended.  If we take the limit $r \rightarrow \infty$,  the electric flux approaches the constant value
\begin{align}
Q_{\rm E}=\frac{4\pi}{\kappa}{\frac {{4m}^{3}\sinh  2\alpha  
  }{ \left(  \cosh^2 \alpha (4m^2-1)+1 \right)^2 }},
\end{align}
which for the typical  choice $\sinh \alpha=1$, $\cosh \alpha =\sqrt{2}$, and $m=1$, takes the constant value
\begin{equation}
Q_{\rm E}=\frac{32 \pi \sqrt{2}}{49 \kappa}.
\end{equation}

The magnetic flux for the boosted Kaluza-Klein magnetic monopole turns out to be 
\begin{align}
\Phi_{{\rm B}}=&\oint_{s^2} \frac{1}{2}\tilde{ F}^{\alpha \beta}{\rm d}s_{\alpha \beta}=\oint_{s^2} \frac{1}{4}\eta^{\alpha \beta \mu \nu}F_{\mu \nu}{\rm d}s_{\alpha \beta}=
\oint_{s^2} \frac{1}{2}|g^{(2)}| g^{tt}g^{rr}g^{\theta \theta}g^{\phi \phi}F_{\theta \phi}{\rm d}\theta{\rm d}\phi,
\end{align}
which gives
\begin{equation}
\Phi_{{\rm B}}=\frac{2\pi}{\kappa}\oint_{s^2}-\frac{r \sin \theta  \cosh \alpha  \left(\sinh ^2\alpha \left(8
   m^2+r \cos 2 \theta  (4 m-r)-4 m r+r^2\right)+8 m^2 r \sin ^2\theta
    (2 m-r) \cosh ^2\alpha\right)^2}{8 (2 m-r) \left(4 m^2 r \sin
   ^2\theta (2 m-r) \cosh ^2\alpha+\sinh ^2\alpha\left(3 m^2
   \cos ^2\theta +\sin ^2\theta  (r-2 m)^2\right)\right)^2}{\rm d}\theta,
\end{equation}
which again,  is a function of $r$. Taking the limit, $r \rightarrow \infty$ one obtains
\begin{equation}
\Phi_{{\rm B}}=\frac{2\pi}{\kappa} \cosh \alpha.
\end{equation}
The electric, and magnetic fluxes for the four dimensional metric deduced from  (\ref{15}) are valid for both signs of $m$.

\begin{center}
\begin{figure} \hspace{4.cm}\includegraphics[width=8.cm]{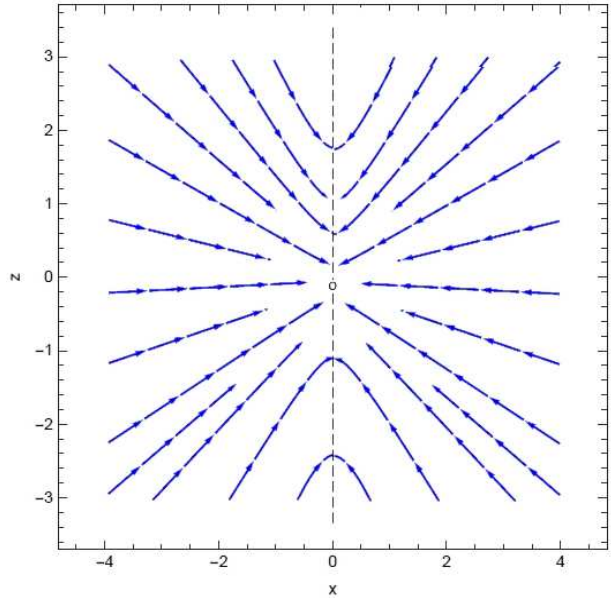}\caption{\label{B} \small
Vector plots for  $B_x$ and $B_z$ in the case with $\cosh \alpha=\sqrt{2}$, $\sinh \alpha =1$, $m=1$ in the x-z plane, which corresponds to a dyon  at the origin plus a magnetically charged string along the $z$ axis.  }
\end{figure}
\end{center}

We conclude that,  the magnetic monopole gives rise to a dyon plus after the boost \cite{kible}. To see this,  we convert the magnetic fields $B_r$ and $B_{\theta}$ from spherical coordinates to a cartesian one on the  $(x,z)$ plane, that is
\begin{eqnarray}
B_{x}&=&B_{r}\sin \theta+B_{\theta}\cos\theta~, \\
B_{z}&=&B_{r}\cos \theta-B_{\theta}\sin\theta,
\end{eqnarray}
using $x=r\sin\theta$, $z=r\cos\theta$. The magnetic fields $B_{x}$ and $B_{z}$ are shown in Fig. (\ref{B}).  The vector field illustrates that  there should be a dyon  at the origin plus a string along the $z$ axis.

For the boosted Kaluza-Klein magnetic monopole, we can obtain the conserved quantities in the spacetime, in the presence of Killing vectors $\xi^{\mu}=\delta ^{\mu}_{t}$ and $\xi^{\mu}=\delta ^{\mu}_{\phi}$, which correspond to time translation and axial symmetry, respectively. To do this, we use the following integral for the conserved quantities \cite{padm}
\begin{equation}
I=\frac{1}{8\pi G}\int_{s}\nabla^{n}\xi^{m} {\rm d}^2\Sigma_{mn},
 \end{equation} 
 where ${\rm d}^2\Sigma_{mn}$ should be taken over a two-dimensional surface located at the spatial infinity, that can be calculated via the metric components (\ref{44}).
 If we consider the time translation invariance $\xi^{\mu}=\delta ^{\mu}_{t}$, and substituting into the integral we get
 \begin{equation}
 M=\frac{m}{2G}\frac{\sinh 2\alpha}{4m^2\cosh^2\alpha-\sinh^2 \alpha},
 \end{equation}
which is the total mass of the system. 
Note  however, that  in the case of axial symmetry, where the relevant Killing vector is $\xi^{\mu}=\delta ^{\mu}_{\phi}$, the integral will give $J=0$, where $J$ is the angular momentum.
 Also note that $M=0$ for $\alpha=0$, in agreement with the results of section \ref{3}. The boost, therefore generates mass and electric charge.

\section{Conclusion}
Inspired by  the Taub-NUT solution, we  considered a Kaluza-Klein
vacuum solution in five dimensions, which described  a point-like   magnetic monopole in
four dimensional  spacetime. The source  supporting the four dimensional
space-time was shown to differ from that of an ultra-relativistic
fluid, in contrast to the solution of  Wesson and Leon \cite{23}. The
pressure is anisotropic in both cases. 
 We calculated the magnetic charge
and showed that the total magnetic flux of the monopole through
any spherical surface centered at the origin was constant,
indicating that  there is no extended
magnetized source. The gravitational mass was derived in two ways
and it was shown to vanish using both definitions.  It
was pointed out that the singularity which appears at finite $r$
is neither a horizon, nor a surface of finite, non-vanishing
surface area. In a more appropriate coordinate system, this was
shown to be a  curvature singularity at ${\tilde r}=0$. 
The main contribution  of this paper was  studying  the properties of the 
 boosted Kaluza-Klein magnetic monopole. It was shown that the boosted solution acquires significantly different physical properties in $(3+1)$ dimensions, including the appearance of a magnetically charged string attached to a dyon  with extended electric and magnetic charges.We considered both signs $\pm$ for $m$ in  our calculations, and  showed that for obtaining the main results of the paper, which appear after boost, 
 it was not essential to choose a particular sign for $m$.
\\

{\bf Acknowledgements}\\

The authors would like to thank the anonymous referee for
helpful comments. N.R. Acknowledges the support of Shahid Beheshti University.


\begin{thebibliography}{99}
\bibitem{RS}
L. Randall and R. Sundrum. Mod. Phys. Lett. A13, 2807 (1998); L. Randall and R. Sundrum.
Phys. Rev. Lett. 83, 4690 (1999).
\bibitem{stm}
 P. S. Wesson, Phys. Lett. B276, 299 (1992); P. S. Wesson, Mod. Phys. Lett. A7, 921 (1992); P. S. Wesson, J. Math. Phys. 33, 3883 (1992); J. Ponce de Leon, P. S. Wesson, J. Math. Phys. 34, 4080 (1993); J. M. Overduin and P. S. Wesson. Phys. Rept 283, 303 (1997); J. Ponce de Leon, Class. Quant. Grav. 23, 3043 (2006); B. Mashhoon, P. S. Wesson, Gen. Rel. Grav. 39, 1403 (2007).
\bibitem{1}
T. Kaluza. Sitzungsber Preuss Akad Wiss. Berlin. (Math. Phys.),
996, (1961); O. Klein, Z. Phys. 37,  895, (1926).
\bibitem{5}
D. J. Gross  and M. J. Perry.  Nucl. Phys. B,  226,  29, (1983).
\bibitem{17}
 P. S. Wesson and J. Ponce de Leon.  Astronomy and Astrophysics 294, 1, (1995).
\bibitem{13}
O. Klein, Zeits. Phys. 37, 895, (1926).
\bibitem{18}
P. S. Wesson and J. Ponce de Leon.  Journal of Math. Phys. 33.11,
3883, (1992).
\bibitem{14}
J. M. Overduin, and Paul S. Wesson. Physics Reports 283.5, 303,
(1997).
\bibitem{16}
J. Ponce de Leon and P. S. Wesson. Journal of Math. Phys. 34.9,
4080-4092, (1993).
\bibitem{8}
G. Lessner, Phys. Rev. D25, 3202, (1982).
\bibitem{Y}
Y. Thiry. Comptes Rendus de la Academie des Sciences (Paris) 226,
216, (1948).
\bibitem{2}
O. Heckmann, P. Jordan  and W. Fricke.  Z. Astrophys. 28, 113,
(1951).
\bibitem{7}
P. S. Wesson,  Phys. Lett, 276B, 299, (1992).
\bibitem{26}
R. D. Sorkin.  Physical Review Letters 51.2,  87, (1983). R. D.
Sorkin,  Phys. Rev. Lett, 51, 87, (1983).
\bibitem{G}
P. Bizoń, T. Chmaj and G. Gibbons.  Physical review letters,  96(23): 231103,  (2006).
\bibitem{j}
Y. Kanou.  Physical Review D 90.8,  084004, (2014).
\bibitem{Dirac}
P. A. Dirac, Proc. Roy. Soc. A133, 60 (1931).
\bibitem{M}
A. Macias, T. Matos.  Classical and Quantum Gravity,  13(3):345, (1996).
\bibitem{ortin}
T. Ortin, Gravity and String, Cambridge University Press, (2004).
\bibitem{UK}
R. Jante R and BJ. Schroers.  Journal of Geometry and Physics, 104:305-28, (2016).
 \bibitem{21}
 D. Baleanu and  S. Codoban. General Relativity and Gravitation, 31(4), 497,  (1999).
 \bibitem{22}
I. Cotăescu.  Physical Review D 72, no. 4, 044007, (2005).
\bibitem{k}
D. A. Konkowski, T. M. Helliwell and  L. C.  Shepley. Physical Review D, 31(6), 1178, (1985). 
 \bibitem{20}
 J. B. Griffiths and J. Podolsky.  Exact Space-Times in Einstein\rq{}s General Relativity, Cambridge University Press, (2009).
 \bibitem{4}
R. M. Wald, General Relativity, Chicago: University of Chicago
Pres, (1984).
\bibitem{28}
M. Alfredo, and T. Matos. Classical and Quantum Gravity 13, no. 3,
345, (1996).
\bibitem{ssh}
N. Riazi, and S. S.  Hashemi.   International Journal of Modern Physics: Conference Series (Vol. 41, p. 1660121). World Scientific Publishing Company  (2016).
 \bibitem{25}
J. W.  Van Holten.  Physics Letters B 342.1  47-52, (1995).
\bibitem{23}
P. S. Wesson and J. Ponce de Leon.  Classical and Quantum
Gravity 11.5, 1341, (1994).
\bibitem{caroll}
Sean M Carroll,  Spacetime and geometry. An introduction to general relativity. Vol. 1. (2004).
\bibitem{kible}
T. W. B. Kibble and T. Vachaspati.  Journal of Physics G: Nuclear and Particle Physics 42.9 094002 (2015).
\bibitem{padm}
T. Padmanabhan,  Gravitation: foundations and frontiers. Cambridge University Press, (2010).
\end{thebibliography}
\end{document}